\documentclass[american,aps,pra,superscriptaddress,twocolumn]{revtex4-2}

\usepackage[T1]{fontenc}
\usepackage[sc]{mathpazo}
\usepackage{amsmath}
\usepackage{amssymb}
\usepackage{enumerate}
\usepackage{amsthm}
\usepackage{color}
\usepackage{graphicx} 

\usepackage{hyperref}
\hypersetup{pdfpagemode=UseNone}

\newcommand{\bra}[1]{\langle#1|}

\newcommand{\ket}[1]{|#1\rangle}

\newcommand{\braket}[1]{\langle#1\rangle}

\begin{document}

\title{ An uncertainty relation in the case of four observables}

 \author{Minyi Huang}
 \email{11335001@zju.edu.cn}
 \email{hmyzd2011@126.com}
 \affiliation{Department of Mathematical Sciences, Zhejiang Sci-Tech University, Hangzhou 310018, PR~China}



\begin{abstract}
Uncertainty is a fundamental and important concept in quantum mechanics. 
In this work, using the technique in matrix theory, we propose an uncertainty relation of four observables and show that 
the uncertainty constant is tight. It is argued that this method can deal with the several known uncertainty relations for two, three and four observables in a unified way. The result is also compared with other uncertainty relations of four observables. 
\end{abstract}

\maketitle

\section{Introduction}
Uncertainty is one of the most fundamental concepts in quantum mechanics, playing an important role in the researches
of quantum information and computation, as well as their various applications.

The history of uncertainty dates back to the early time of quantum mechanics.
In 1927, Heisenberg discovered his profounding uncertainty principle which becomes a fundamental relation in quantum mechanics,
showing the precision limitation of joint measurement for incompatible observables \cite{heisenberg1927anschaulichen}. In 1929, the Robertson uncertainty
relation is formulated through the variances of two observables \cite{robertson1929uncertainty}.
For any two quantum mechanical observables $A$ and $B$, the Robertson uncertainty can be written as
\begin{equation}
\Delta A\Delta B\geqslant \frac{1}{2}|\braket{A,B}|, \label{uncertain}
\end{equation}
where $\braket{\Omega}$ is the expected value of the observable $\Omega$ with respect to the state $\rho$,
$(\Delta \Omega)^2\triangleq \braket{\Omega^2}-\braket{\Omega}^2$ and
$\Delta \Omega\triangleq \sqrt{\braket{\Omega^2}-\braket{\Omega}^2}$ are the variance and standard deviation, respectively.
Since then, lots of discussions are devoted to the generalization of the uncertainty relation,  varying from more observables to different metrics such as entropy \cite{robertson1934indeterminacy,maassen1988generalized,arthurs1988quantum,ozawa2004uncertainty,busch2007heisenberg,
wehner2008higher,wehner2010entropic,watanabe2011uncertainty,huang2012variance,chen2015sum,song2016stronger,chen2016variance,li2016equivalence,coles2017entropic,
miyadera2008heisenberg,dodonov2018variance,dodonov2019uncertainty,chen2020observable}. The discussions of uncertainty has stimulated the studies of quantum nonlocality, quantum entanglement and many other topics~\cite{horodecki2009quantum}.  Meanwhile, it is also an ingredient of many problems
in mathematical physics~\cite{bell1964einstein,cirel1980quantum,landau1987violation}.

 In the case of three observables,  the tightest uncertainty constant in the summation form was discussed in ~\cite{song2016stronger}. Recently, by introducing an ancillary system and using the properties of the Pauli operators, an investigation reveals the tightest uncertainty constants for both product and summation forms~\cite{liang2024signifying}. It is showed that
 \begin{eqnarray}
&&\prod_{j=1}^{3}(\Delta H_j)^2\geqslant (\frac{1}{\sqrt{3}})^3\prod_{j=1}^{3}|\braket{[H_j,H_{j+1}]}|,\label{varpro}\\
&&\sum_{j=1}^{3}(\Delta H_j)^2\geqslant\frac{1}{\sqrt{3}}\sum_{j=1}^{3}|\braket{[H_j,H_{j+1}]}|. \label{var2}
\end{eqnarray}


In this paper, we propose an uncertainty relation of four observables. The structure of the paper is organized as follows.
In Section II, we prove the uncertainty relation of four observables and the corresponding uncertainty constant is obtained. Section III shows that one can investigate the mentioned uncertainty relations for two, three and four observables in a unified framework. Section IV shows the tightness of the the constant of four observables. The result is also compared with other uncertainty relations 
of four observables. 
Section V is some discussions and the summary.

\section{The uncertainty relations and constants with four observables}
Let $H_j, (j=1,2,3,4)$ be four Hermitian operators and consider the following operator
\begin{eqnarray}
R=\begin{bmatrix}
H_1+ i H_2& i H_3+H_4\\
i H_3-H_4 & H_1- i H_2\\
\end{bmatrix}.\label{4}
\end{eqnarray}
Direct calculations show that
\begin{eqnarray*}
\nonumber RR^\dag=\begin{bmatrix}
R_1&R_2\\
R_3&R_4 \\
\end{bmatrix},
\end{eqnarray*}
where
\begin{eqnarray}
\nonumber R_1&=&\sum_{j=1}^4 H_j^2- i[H_1, H_2]+i[H_3,H_4],\\
\nonumber R_2&=&-i[H_1, H_3]-i[H_2,H_4]-[H_1,H_4]+[H_2, H_3],\\
\nonumber R_3&=&-i[H_1, H_3]-i[H_2,H_4]+[H_1,H_4]-[H_2, H_3],\\
\nonumber R_4&=&\sum_{j=1}^4 H_j^2+i[H_1, H_2]-i[H_3,H_4].
\end{eqnarray}

Now since $RR^\dag$ is semi-definite positive, then for any semi-definite positive operator $\rho$, we have
$Tr (\rho RR^\dag)\geqslant 0$. Moreover, assume that
\[
\rho=\begin{bmatrix}
\cos^2\alpha \nu & r\nu \\
r^*\nu & \sin^2\alpha \nu \\
\end{bmatrix},
\]
where $\nu$ is an state and $r$ is some complex number such that $|r|^2\leqslant \cos^2\alpha\sin^2\alpha$.
Note that this ensures that $\rho\geqslant 0$. In fact, one can denote
$\mu=\begin{bmatrix}
\cos^2\alpha & r\\
r^* & \sin^2\alpha \\
\end{bmatrix}$. Then $\rho=\mu\otimes\nu$, where $\mu$ is a state since $|r|^2\leqslant \cos^2\alpha\sin^2\alpha$.

Now direct calculations show that
\begin{small}
\begin{eqnarray}
\nonumber &&Tr (\rho RR^\dag)\\
\nonumber &=&\cos^2\alpha Tr\nu(\sum_{j=1}^4 H_j^2- i[H_1, H_2]+i[H_3,H_4])\\
\nonumber &+&r Tr \nu(-i[H_1, H_3]-i[H_2,H_4]+[H_1,H_4]-[H_2, H_3])\\
\nonumber &+&r^* Tr \nu(-i[H_1, H_3]-i[H_2,H_4]-[H_1,H_4]+[H_2, H_3])\\
\nonumber &+&\sin^2\alpha Tr\nu(\sum_{j=1}^4 H_j^2+ i[H_1, H_2]-i[H_3,H_4])\\
\nonumber &=& Tr\nu(\sum_{j=1}^4 H_j^2)+(i\sin^2\alpha-i\cos^2\alpha)Tr\nu ([H_1, H_2]-[H_3,H_4])\\
\nonumber &+&2 Re [r Tr\nu (-i[H_1, H_3]-i[H_2,H_4]+[H_1,H_4]-[H_2, H_3])].\\
\label{5}
\end{eqnarray}
\end{small}

Note that since $Tr\nu(-i[H_1, H_3]-i[H_2,H_4])$ is real and $Tr\nu([H_1,H_4]-[H_2, H_3])$ is imaginary, by the property of complex numbers (or equivalently the one dimensional Schwarz inequality for complex numbers),
\begin{small}
\begin{eqnarray}
\nonumber&& |2 Re [r Tr\nu (-i[H_1, H_3]-i[H_2,H_4]+[H_1,H_4]-[H_2, H_3])]|\\
\nonumber&\leqslant& 2|r|\sqrt{|\braket{[H_1, H_3]}+\braket{[H_2,H_4]}|^2+|\braket{[H_1,H_4]}-\braket{[H_2, H_3]}|^2}.\\
\label{6}
\end{eqnarray}
\end{small}
Moreover, one can choose appropriate $\alpha$ and $r$ such that
\begin{small}
\begin{eqnarray}
\nonumber&& (i\sin^2\alpha-i\cos^2\alpha)Tr\nu ([H_1, H_2]-[H_3,H_4])\leqslant 0,\\
\nonumber&& 2 Re [r Tr\nu (-i[H_1, H_3]-i[H_2,H_4]+[H_1,H_4]-[H_2, H_3])]\leqslant 0.
\end{eqnarray}
\end{small}
Thus according to the Schwarz inequality, one can see that Eq. (\ref{5}) reduces to
\begin{small}
\begin{eqnarray}
\nonumber &&Tr (\rho RR^\dag)\\
\nonumber &\geqslant& Tr\nu(\sum_{j=1}^4 H_j^2) -\sqrt{(\sin^2\alpha-\cos^2\alpha)^2+4|r|^2}\\
\nonumber &&\sqrt{|\braket{[H_1, H_2]}-\braket{[H_3,H_4]}|^2+|\braket{[H_1, H_3]}}\\
\nonumber &&\overline{+\braket{[H_2,H_4]}|^2+|\braket{[H_1,H_4]}-\braket{[H_2, H_3]}|^2}
\end{eqnarray}
\end{small}
Note that
\begin{eqnarray}
\nonumber&&(\sin^2\alpha-\cos^2\alpha)^2+4|r|^2\\
\nonumber&\leqslant& (\sin^2\alpha-\cos^2\alpha)^2 +4\cos^2\alpha\sin^2\alpha\\
&=&1.\label{7}
\end{eqnarray}
Combining Eqs. (\ref{6}) and (\ref{7}), one has
\begin{small}
\begin{eqnarray}
\nonumber &&Tr (\rho RR^\dag)\\
\nonumber &\geqslant& Tr\nu(\sum_{j=1}^4 H_j^2) -\sqrt{|\braket{[H_1, H_2]}-\braket{[H_3,H_4]}|^2+|\braket{[H_1, H_3]}}\\
 &&\overline{+\braket{[H_2,H_4]}|^2+|\braket{[H_1,H_4]}-\braket{[H_2, H_3]}|^2}.\label{8}
\end{eqnarray}
\end{small}
For simplicity, the above equation can be rewritten as
\begin{small}
\begin{eqnarray}
\nonumber &&Tr (\rho RR^\dag)\geqslant Tr\nu(\sum_{j=1}^4 H_j^2)-\\
 && \sqrt{\sum_{i<j,k<l}|\braket{[H_i, H_j]}-(-1)^{\tau(ijkl)}\braket{[H_k,H_l]}|^2},\label{20}
\end{eqnarray}
\end{small}
\noindent where $(ijkl)$ is a permutation of $(1234)$ and $\tau(ijkl)$ is the number of inversions.

One can discuss when the identity in Eq. (\ref{20}) holds. Apparently, the identity in Eq. (\ref{20}) holds
 iff the Schwarz identities in Eqs. (\ref{6}) and (\ref{7}) hold. However, this can be easily done by
 by simply choosing appropriate $\alpha$ and $r$.
 Thus for any state $\nu$ and the specified $\rho=\mu\otimes\nu$, we have
 \begin{eqnarray}
\nonumber &&Tr (\rho RR^\dag)= Tr\nu(\sum_{j=1}^4 H_j^2)-\\
 && \sqrt{\sum_{i<j,k<l}|\braket{[H_i, H_j]}-(-1)^{\tau(ijkl)}\braket{[H_k,H_l]}|^2}.
\end{eqnarray}
Now by $Tr (\rho RR^\dag)\geqslant 0$ and the quadratic mean inequality, we have
\begin{small}
\begin{eqnarray*}
Tr\nu(\sum_{j=1}^4 H_j^2)\geqslant \frac{1}{\sqrt{3}}\sum_{i<j,k<l}|\braket{[H_i, H_j]}-(-1)^{\tau(ijkl)}\braket{[H_k,H_l]}|.
\end{eqnarray*}
\end{small}
By replacing $H_j$ with $H_j-\braket{H_j}I$, one can obtain
\begin{small}
\begin{eqnarray}
\nonumber\sum_{j=1}^4 (\Delta H_j)^2\geqslant \frac{1}{\sqrt{3}}\sum_{i<j,k<l}|\braket{[H_i, H_j]}-(-1)^{\tau(ijkl)}\braket{[H_k,H_l]}|.\\
\label{4observables}
\end{eqnarray}
\end{small}
Thus we obtain an uncertainty relation in the case of four observables, which is lower bounded by the differences of their commutators.

%

\section{A unified way to discuss the uncertainty relations for two, three and four observables}

It should be mentioned that the operator $R$ can be used to show the Robertson uncertainty relation.  If one sets $H_3$ and
$H_4$ to be zero in Eq. (\ref{4}) and the state $\rho=\ket{0}\bra{0}\otimes \nu$, then $Tr(\rho RR^\dag)$ reduces to $Tr \nu (H_1^2+H_2^2)\geqslant \braket{i[H_1,H_2]} $. Similarly, taking $\rho=\ket{1}\bra{1}\otimes \nu$, one can see that
$Tr \nu (H_1^2+H_2^2)\geqslant -\braket{i[H_1,H_2]}$. Thus
\begin{equation}
Tr [\nu (H_1^2+H_2^2)]\geqslant |\braket{i[H_1,H_2]}|.
\end{equation}
By replacing $H_j$ with $H_j-\braket{H_j}I$, we see that
\begin{equation}
(\Delta H_1^2+\Delta H_2^2)\geqslant |\braket{i[H_1,H_2]}|.
\end{equation}
If $Tr \nu \Delta H_1^2=Tr \nu \Delta H_2^2$, then the above equation reduces to
\begin{equation}
\Delta H_1\Delta H_2\geqslant \frac{1}{2}|\braket{i[H_1,H_2]}|.\label{Ro}
\end{equation}
If If $\Delta H_1^2\neq \Delta H_2^2$, then one can take $\kappa_j=\frac{\sqrt{\Delta H_1\Delta H_2}}{\Delta H_j}$ and
$H_j'=\kappa_j H_j$. One can verify that $\Delta H_1\Delta H_2=\Delta H_1'\Delta H_2'$ and $\braket{i[H_1,H_2]}=\braket{i[H_1',H_2']}$. However, since $\Delta H_1'=\Delta H_2'$, one can see that Eq. (\ref{Ro}) is valid for $H_j'$. By the previous discussions, it is also valid for any observable $H_j$, which is just the Robertson uncertainty relation.

If we take $H_4=0$ in Eq. (\ref{4}), then one can see immediately that it reduces to Eq. (\ref{var2}), i.e. the uncertainty relation of three observables. Moreover, according to Ref. \cite{huang2025triple}, Eq. (\ref{varpro}) can be obtained as a corollary of Eq. (\ref{var2}). 
Thus our result gives a unified way to derive the uncertainty relations for two, three and four observables.

\section{The tightness of the uncertainty constant and the comparison with other uncertainty relations}
It should also be pointed out that the constant $\frac{1}{\sqrt{3}}$ is tight in the case of four observables. Note that if we take $H_4=I$, then Eq. (\ref{4observables}) reduces to Eq. (\ref{var2}), i.e. the summation form of three observables.
However, in Ref. \cite{liang2024signifying},
$\frac{1}{\sqrt{3}}$ is proved to be tight. Thus it can also be achieved in the case of four observables.

On the other hand, one may wonder whether the there are nontrivial examples for which all the $H_j$ are noncommutative such that the identity in Eq. (\ref{4observables}). Indeed, this is the case.
To see this, let us take
$H_j=\begin{bmatrix}
1&e^{-i\beta_j}\\
e^{i\beta_j}&1
\end{bmatrix}$ for $j=2,3,4$, where $|\beta_2-\beta_3|=|\beta_2-\beta_4|=|\beta_3-\beta_4|=\frac{2\pi}{3}.$ Take
$H_1=\begin{bmatrix}
\lambda_1&0\\
0&\lambda_2
\end{bmatrix}$, where $\lambda_i$ are real constants. By choosing the state to be $\ket{0}\bra{0}$, one can see through direct calculations that the identity holds in Eq. (\ref{4observables}). \\

One can compare the result with that in Ref. \cite{liang2024signifying}. In \cite{liang2024signifying}, the uncertainty relation is
\begin{eqnarray*}
\sum_{j=1}^m (\Delta H_j)^2\geqslant r_m\sum_{i<j}^m\frac{2|\braket{[H_i, H_j]}|}{m-1},
\label{m observables2}
\end{eqnarray*}
where $r_m= \inf \{ \frac{(m-1)\sum_{j=1}^m (\Delta H_j)^2}{\sum_{i<j}^m 2|\braket{[H_i, H_j]}|}\} $. Since all the $|\braket{[H_i, H_j]}|$ has the same coefficients, the above uncertainty relation is equivalent to
\begin{eqnarray}
\sum_{j=1}^m (\Delta H_j)^2\geqslant k_m\sum_{i<j}^m |\braket{[H_i, H_j]}|,
\label{m observables3}
\end{eqnarray}
where $k_m= \inf \{ \frac{\sum_{j=1}^m (\Delta H_j)^2}{\sum_{i<j}^m|\braket{[H_i, H_j]}|}\}$. However, by choosing $H_1=I$ and
$H_j=\begin{bmatrix}
1&e^{-i\beta_j}\\
e^{i\beta_j}&1
\end{bmatrix}$ for $j=2,3,4$, where $|\beta_2-\beta_3|=|\beta_2-\beta_4|=|\beta_3-\beta_4|=\frac{2\pi}{3}$ as above, we see immediately that $k_m\leqslant\frac{1}{\sqrt{3}}$. Moreover, if we take $H_1=\begin{bmatrix}
d&c\\
c^*&d
\end{bmatrix}$ and the same $H_j (j=2,3,4)$ as above, then direct calculations show that
\begin{equation}
\sum_{j=2}^{4}(\Delta H_j)^2=\frac{1}{\sqrt{3}}\sum_{j=2}^{4}|\braket{[H_j,H_{j+1}]}|.\label{16}
\end{equation}
On the other hand,
\begin{eqnarray*}
&&(\Delta H_1)^2=|c|^2\\
&&\sum_{j=2}^4 |\braket{[H_1,H_j]}|=\sum_{j=2}^4 2 Im (ce^{-i\beta_j}).
\end{eqnarray*}
Apparently, when $c$ is sufficiently small,
\begin{equation}
(\Delta H_1)^2<\frac{1}{\sqrt{3}}\sum_{j=2}^4 |\braket{[H_1,H_j]}|.\label{17}
\end{equation}
Combining Eqs. (\ref{16}) and (\ref{17}), one can see immediately that
\[
k_4<\frac{1}{\sqrt{3}}.
\]
Thus when $H_1=I$, we see that the uncertainty relation is strictly tighter than that in Eq. (\ref{m observables3}), i.e. the uncertainty relation in
\cite{liang2024signifying}.

\begin{figure}
	\centering
	\includegraphics[width=80 mm]{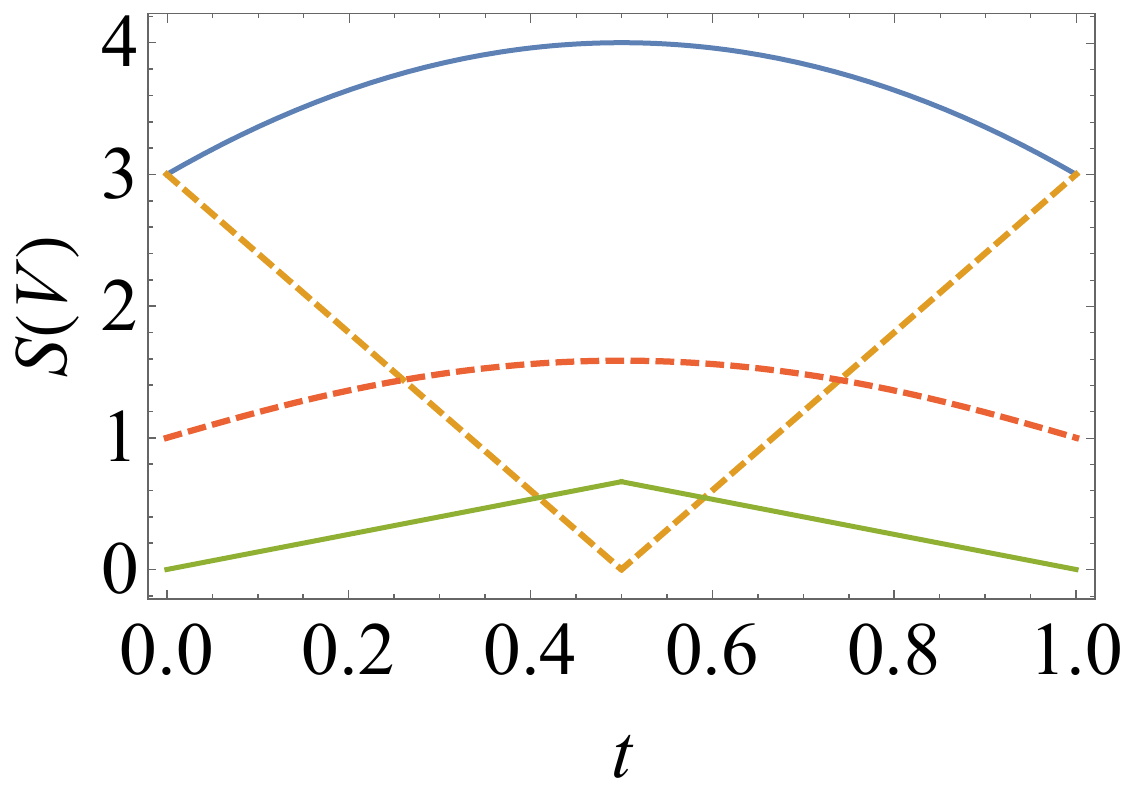}
	\caption{The figure of the The blue solid line is the sum of the variances,
    the brown line is the bound in this paper. Apparently, when $t=0,1$ the values of the brown and the blue curves are the same, showing the tightness of the constant in this paper. 
     The red and green lines are the bounds in Refs. \cite{chen2015sum,chen2016variance}, which gives strictly smaller values than the blue curve. }
    \label{2}
\end{figure}

One can also compare the result with those in Refs. \cite{chen2015sum,chen2016variance}. In Ref. (\ref{chen2015sum}), the uncertainty relation is 
\begin{eqnarray}
\sum_{j=1}^n (\Delta H_j)^2\geqslant\frac{1}{\lambda_{max}(M)}[\Delta(\sum_{j=1}^n H_j)]^2,
\end{eqnarray}
where $M$ is a matrix whose entries are defined by $M_{ij}=Tr [\frac{(H_i-\braket{H_i})\sqrt{\nu}}{\Delta H_i} \frac{(H_j-\braket{H_j})\sqrt{\nu}}{\Delta H_j}]$ and $\lambda_{max}$ is the maximal eigenvalue of $M$.
In Ref. \cite{chen2016variance}, the uncertainty relation is given by 
\begin{eqnarray*}
\sum_{j=1}^n (\Delta H_j)^2\geqslant\frac{1}{n-2}\{\sum_{1\leqslant i<j\leqslant n}^n[\Delta(H_i+H_j)]^2\\
-\frac{1}{(n-1)^2}[\sum_{1\leqslant i<j\leqslant n}^n\Delta(H_i+H_j)]^2\}.
\end{eqnarray*}
Now take $n=4$, the state
$\nu=\begin{bmatrix} 1-t & 0 \\ 0& t\end{bmatrix} (0\leqslant t\leqslant 1)$ and $H_1=\begin{bmatrix}
1&0\\
0&-1
\end{bmatrix}$ $H_j=\begin{bmatrix}
1&e^{-i\beta_j}\\
e^{i\beta_j}&1
\end{bmatrix}$ for $j=2,3,4$, where $\beta_2=\frac{2\pi}{3}, \beta_3=\frac{4\pi}{3}, \beta_4=2\pi.$ Now direct calculations
show that $\sum_{j=1}^4 (\Delta_j)^2=3+4t-4t^2$, $\frac{1}{\sqrt{3}}\sum_{i<j,k<l}|\braket{[H_i, H_j]}-(-1)^{\tau(ijkl)}\braket{[H_k,H_l]}|=3|(1-2t)|$, $\Delta (L_1+L_2)=\Delta (L_1+L_3)=\Delta (L_1+L_4)=1+4t-4t^2$, $\Delta (L_2+L_3)=\Delta (L_2+L_4)=\Delta (L_3+L_4)=1$, $\lambda_{max}=Max\{3-3t,3t\}$ and $\Delta (L_1+L_2+L_3+L_4)=4t(1-t)$. It
is shown in Fig. \ref{2} that for a wide range of $t$, our uncertainty relation is stronger
than those in Refs. \cite{chen2015sum,chen2016variance}.

\section{Discussions}
As mentioned in section III, the result in this paper gives a unified way to derive the uncertainty relation for two, three and four observables. In particular, the result can reduce to the uncertainty relation of three observables. Hence it also gives another proof of those uncertainty relations. 
Comparison can be made between the different proofs.  The derivation in Ref. \cite{liang2024signifying}, the proof introduces an ancillary system and depends on the properties of Pauli operators. It should also be noted that in Ref.~\cite{dodonov2018variance}, Dirac operators are used to discuss variance uncertainty relations. Unlike the ways in \cite{dodonov2018variance,liang2024signifying},
 our approach relies on the techniques of matrix theory without directly introducing such noncommutative operators and their properties but concerns with two scalar parameters $\alpha$ and $r$. Such a way is more analogous to Robertson's original proof of his uncertainty relation, and leads to a large simplification as well as a clear intuition in the concrete calculations and derivations.

 However, it is still possible to consider the connections between the two approaches. In fact, if one notices that the operator $R$ in Eq. (\ref{4}) can
 be rewritten as $R=I\otimes H_1+i\sigma_3\otimes H_2+i\sigma_1\otimes H_3$, then the operator $RR^\dag$ can be rewritten as
 $\sum_j H_j^2-i\sigma_3\otimes[H_1,H_2]-i\sigma_1\otimes[H_1,H_3]+i\sigma_2\otimes[H_2,H_3]$. That is, although we consider the
 problem from the perspective of matrix theory rather than the Pauli operators, the result is implicitly related to them.
 On the other hand, in Ref. \cite{liang2024signifying}, the observable is $V=\sum_{j=1}^3 \frac{H_j\otimes \sigma_j}{\Delta H_j}$ and
 $V^2=3-\sum_{j=1}^3\frac{i\braket{H_j, H_{j+1}}\braket{\sigma_{j+2}}}{\Delta H_j \Delta H_{j+1}}$.
 One can find that both the two seemingly different approaches utilizes the anti-commutativity of Pauli operators to ensure the existence of the commutators of $H_j$. However, this paper shows that when concerning with three observables, only two Pauli operators is enough for the operator $R$. In particular, in the later derivation of the uncertainty constant, one only needs two scaler parameters to determine such a bound. In the case of four observables, the operator $R$ can be written as $R=I\otimes H_1+i\sigma_3\otimes H_2+i\sigma_1\otimes H_3+i\sigma_2\otimes H_4$. With one more observable, one more Pauli operator is needed.
 Interestingly, it does not change the uncertainty constant and the the uncertainty relation still depends on the two scalar parameters, rather than some noncommutative operators. As shown in Eq. (\ref{4observables}), in some sense, the effect a new Pauli operator is reflected by the differences between commutators.
A reduction not to the noncommutative Pauli operators but to two scalar parameters largely simplifies the derivations and naturally generalizes the discussions to the case of four observables.

Thus, our work does not only obtain an uncertainty relation but also gives a unified framework to derive the uncertainty relations for two, three and four observables.


\section*{Acknowledgement}
This work is partially supported by the National Natural Science Foundation of China (12371135).

\end{document}